\documentclass[conference]{IEEEtran}
\usepackage{cite}
\usepackage{amsmath,amssymb,amsfonts}
\usepackage{algorithmic}
\usepackage{graphicx}
\usepackage{booktabs}
\usepackage{textcomp}
\usepackage{xcolor}
\usepackage{epsfig}
\usepackage{hyperref}
\usepackage{bm}
\usepackage[nolist]{acronym}
\usepackage[caption=false,font=footnotesize]{subfig}
\hypersetup{hidelinks,
	colorlinks=true,
	allcolors=black,
	pdfstartview=Fit,
	breaklinks=true}
\def\BibTeX{{\rm B\kern-.05em{\sc i\kern-.025em b}\kern-.08em
    T\kern-.1667em\lower.7ex\hbox{E}\kern-.125emX}}

\newcommand*{\affmark}[1][*]{\textsuperscript{#1}}

\DeclareMathOperator*{\argmax}{arg\,max}

\newcommand{\PM}{{\rm PM}}

\newcommand{\minus}{\!-\!}
\newcommand{\plus}{\!+\!}
\newcommand{\Rmnum}[1]{\uppercase\expandafter{\romannumeral #1}}

\begin{document}

\title{Soft-Output Fast Successive-Cancellation List Decoder for Polar Codes}

\author{
	\IEEEauthorblockN{Li Shen\affmark[1], Yongpeng Wu\affmark[1], Yin Xu\affmark[1], Xiaohu You\affmark[2], Xiqi Gao\affmark[2], and Wenjun Zhang\affmark[1]}\\
	\IEEEauthorblockA{
	\affmark[1]Department of Electronic Engineering, Shanghai Jiao Tong University, Shanghai, China \\
	\affmark[2]National Mobile Communications Research Laboratory, Southeast University, Nanjing, China \\
	\textit{E-mail:} \{shen-l, yongpeng.wu, xuyin, zhangwenjun\}@sjtu.edu.cn, \{xhyu, xqgao\}@seu.edu.cn
	}
}

\begin{acronym}
	\acro{APP}{a-posteriori probability}
	\acro{AWGN}{additive white Gaussian noise}
	\acro{BER}{bit error rate}
	\acro{BLER}{block error rate}
	\acro{BP}{belief propagation}
	\acro{BPSK}{binary phase shift keying}
	\acro{CRC}{cyclic redundancy check}
	\acro{DMC}{discrete memoryless channel}
	\acro{FSCL}{fast SCL}
	\acro{GLDPC}{generalized low-density parity-check}
	\acro{G-REP}{generalized REP}
	\acro{G-PC}{generalized parity-check}
	\acro{LLR}{log-likelihood ratio}
	\acro{MIMO}{multiple-input multiple-output}
	\acro{ML}{maximum-likelihood}
	\acro{PM}{path metric}
	\acro{QPSK}{quadrature phase shift keying}
	\acro{Rate0}{rate-zero}
	\acro{Rate1}{rate-one}
	\acro{REP}{repetition}
	\acro{SC}{successive cancellation}
	\acro{SCAN}{soft cancellation}
	\acro{SCL}{successive cancellation list}
	\acro{SO-SCL}{soft-output SCL}
	\acro{SO-FSCL}{soft-output FSCL}
	\acro{SO-FSCDL}{SO-FSCL with dynamic list size}
	\acro{SPC}{single-parity-check}
\end{acronym}

\maketitle

\begin{abstract}
The soft-output successive cancellation list (SO-SCL) decoder provides a methodology for estimating the a-posteriori probability log-likelihood ratios by only leveraging the conventional SCL decoder for polar codes. However, the sequential nature of SCL decoding leads to a high decoding latency for the SO-SCL decoder. In this paper, we propose a soft-output fast SCL (SO-FSCL) decoder by incorporating node-based fast decoding into the SO-SCL framework. Simulation results demonstrate that the proposed SO-FSCL decoder significantly reduces the decoding latency without loss of performance compared with the SO-SCL decoder.
\end{abstract}


\section{Introduction}
Arıkan's invention, polar codes, represents an advanced channel coding technique that utilizes the principle of channel polarization \cite{Arikan2009Channel}. This class of channel coding is distinguished by their structured code construction and manageable complexity. With \ac{SC} decoding algorithm \cite{Arikan2009Channel}, polar codes are proven to approach the symmetric capacity of binary-input discrete memoryless channels as the code length tends to infinity. However, in practical scenarios with moderate-to-short code lengths, the effect of polarization may be inadequate, resulting in a performance gap compared to \ac{ML} decoding \cite{Tal2015List}. To mitigate this, the \ac{SCL} decoder \cite{Tal2015List} provides a list of the most likely candidate codewords. By further integrating \ac{CRC} to identify the correct candidate codeword, the \ac{CRC}-aided \ac{SCL} decoding can approach the performance of \ac{ML} decoding \cite{Tal2015List,Niu2012CRC,Balats2015LLR}.

Nevertheless, the sequential nature of \ac{SC} and \ac{SCL} decoders results in high decoding latency, which is difficult to further reduce. Since polar codes can be decomposed into the polarization of two sub-polar codes recursively, many works consider identifying some special subcodes and directly obtaining the estimated codewords of these subcodes \cite{Alamdar2011Simp,Sarkis2014Fast,Hanif2017Fast,Condo2018Generalized,Zheng2021Threshold,Sarkis2016Fast,Hashemi2017Fast,Ardakani2019Fast,Ren2022Sequence}. 

Specifically, \ac{Rate0} and \ac{Rate1} nodes were first identified for \ac{SC} decoder in \cite{Alamdar2011Simp}. Later, \ac{SPC} nodes, \ac{REP} nodes, and some of their combinations were considered in \cite{Sarkis2014Fast}. In addition, more general nodes have been investigated in \cite{Hanif2017Fast,Condo2018Generalized,Zheng2021Threshold} for \ac{SC} decoder. These special nodes are also suitable for SCL decoding \cite{Sarkis2016Fast,Hashemi2017Fast,Ardakani2019Fast,Ren2022Sequence}, with the path splitting and path selection underneath these nodes handled.

In many scenarios, such as \ac{MIMO} systems and bit-interleaved coded modulation systems, a soft-output decoder is required to enable iterative detection and iterative decoding \cite{Hochwald2003Ach, Li2002Bit}. Yet, the above polar decoders are hard-output and fail to provide post-decoding soft information. Following the BCJR algorithm \cite{Bahl1974Optimal}, we can obtain an optimal estimate of the \ac{APP} at the cost of exponential complexity. Some polar soft decoders based on \ac{BP} decoding \cite{Arikan2008Perform} or \ac{SCAN} decoding \cite{Fayyaz2014Low} require iterations or an additional cascaded \ac{SCL} decoder \cite{Yuan2014Early,Elkelesh2018Belief,Pillet2020SCAN,Xiang2020Soft,Egilmez2022soft,Fominykh2023Effic}. For short block length codes, \cite{Yuan2023Soft} presents a universal soft-output decoder that is not only applicable to polar codes. Given a list of candidate codewords, Pyndiah's approximation \cite{Pyndiah1998Near} can provide estimates of the \ac{APP} \acp{LLR}. However, a limited list size may result in infinite values of the approximation that need to be bounded to a saturated value. In \cite{Yuan2024Near}, the proposed \ac{SO-SCL} decoder outputs more accurate estimates by modifying Pyndiah's approximation with a term called codebook probability, leveraging the \ac{SCL} decoding tree.

In this paper, we investigate the fast decoding of \ac{SO-SCL} to reduce the decoding latency and propose a \ac{SO-FSCL} decoder by identifying some special nodes. Since the estimate of codebook probability requires accessing all roots of unvisited subtrees in the \ac{SCL} decoding tree, while the node-based fast decoding may only visit some of them, we need to address this for our \ac{SO-FSCL} decoder. Furthermore, to satisfy the requirement of dynamic frozen bits for codebook probability estimation, we also consider the compatibility of the proposed \ac{SO-FSCL} decoder with dynamic frozen bits.

\section{Preliminaries}
\subsection{Notations}
Random variables are denoted by uppercase letters, e.g., $X$, and their realizations are denoted by corresponding lowercase letters, e.g., $x$. A vector of length $N$ is denoted as $\bm{x}^N = (x_1, x_2, \cdots, x_N)$, where $x_i$ is the $i$-th entry. We denote $p_X$ and $P_Y$ by the probability density function of a continuous random variable $X$ and the probability mass function of a discrete random variable $Y$, respectively. Sets like alphabet are denoted by calligraphic letters, e.g., $\mathcal{X}$. $\mathcal{X}^C$ and $|\mathcal{X}|$ represent the complement and cardinality of $\mathcal{X}$, respectively. An index set $\{i, i+1, \cdots,j\}$ ($j>i$) is abbreviated as $[\![i,j]\!]$, and $[\![1,j]\!]$ is further abbreviated as $[\![j]\!]$. Given a vector $\bm{x}^N$ and $\mathcal{A} \subseteq [\![N]\!]$, we write $\bm{x}_\mathcal{A}$ to denote the subvector $[x_i]$ with all $i\in\mathcal{A}$.

\subsection{Polar Codes}
Assume that a binary polar code $(N,K)$ is of code length $N=2^n$ and code dimension $K$, where $n$ is a positive integer. Among the $N$ polarized subchannels, the $K$ most reliable subchannels are indexed by $\mathcal{I} \subseteq [\![N]\!]$, while the remaining positions are denoted by $\mathcal{F} = [\![N]\!] \cap \mathcal{I}^C$. Thus, the input vector $\bm{u}^N$ for polar transform consists of $\bm{u}_\mathcal{I}$ and $\bm{u}_\mathcal{F}$ that are placed with information bits and frozen bits, respectively. Each frozen bit $u_i$, $i\in\mathcal{F}$, is either set to a static value like zero, or determined as a linear function of previous input $\bm{u}_{[\![i-1]\!]}$, which is also known as the \textit{dynamic frozen bit}. The polar codeword $\bm{c}^N$ is generated by
\begin{equation}
	\bm{c}^N = \bm{u}^N \bm{G}_N,
\end{equation}
where $\bm{G}_N = \bm{G}^{\otimes n}$ is the $n$-th Kronecker power of $\bm{G}=\left[\begin{smallmatrix}1&0\\1&1\end{smallmatrix}\right]$. Then, $\bm{c}^N$ is modulated to \ac{BPSK} symbols, and transmitted over $N$ independent uses of a binary-input \ac{DMC} or \ac{AWGN} channel.

\subsection{SC and SCL Decoding}
At the receiver, the \ac{SC} decoder operates by processing bits in $\bm{u}^N$ sequentially, making decisions on each bit based on the channel observation $\bm{y}^N$ and previously determined bits. In particular, the $i$-th input bit $u_i$ is estimated according to \cite{Arikan2009Channel}
\begin{equation}
    \hat{u}_i=\left\{\begin{aligned}
    &\,\text{frozen value}, && \hspace{-0.1cm} i\in\mathcal{F}; \\
    &\argmax_{u_i\in\{0,1\}} P_{\bm{Y}^N, \bm{U}^{i-1}|U_i}\left(\bm{y}^N, \hat{\bm{u}}_{[\![i-1]\!]}|u_i\right), && \hspace{-0.1cm} i\in\mathcal{I}.
    \end{aligned}\right.
\end{equation}

Unlike the \ac{SC} decoder which just retains the most probable information bit at each decision, the \ac{SCL} decoder considers each information bit being both 0 and 1. Thus, given a list size $L$, the candidate codewords (paths) doubles at each decision on $u_i$ for $i\in\mathcal{I}$, and only $L$ paths with the lowest \acp{PM} survive. After the $i$-th bit decision, the \ac{PM} associated with the $l$-th path, denoted by $\PM^{(l)}_{i}$, is calculated by \cite{Balats2015LLR}
\begin{equation}
	\PM^{(l)}_{i} = \sum_{k=1}^i \ln\left(1 + e^{-(1-2\hat{u}^{(l)}_{k})\lambda_{k}^{(l)}}\right),
	\label{eq:PM_u}
\end{equation}
where $\hat{\bm{u}}^{(l)}$ is the estimated input vector at the $l$-th path and the \ac{LLR} $\lambda^{(l)}_{k}$ is defined by
\begin{equation}
	\lambda^{(l)}_{k} = \ln \frac{P_{\bm{Y}^N, \bm{U}^{k-1} | U^{(l)}_{k}}\left(\bm{y}^N, \hat{\bm{u}}^{(l)}_{[\![k-1]\!]} \big| 0\right)}{P_{\bm{Y}^N, \bm{U}^{k-1} | U^{(l)}_{k}}\left(\bm{y}^N, \hat{\bm{u}}^{(l)}_{[\![k-1]\!]} \big| 1\right)}.
\end{equation}

\subsection{SO-SCL Decoding}
The calculation of bit-wise \ac{APP} \acp{LLR} requires the assistance of a quantity called the \textit{codebook probability} in \cite{Yuan2024Near}, which is written as
\begin{equation}
	P_{\mathcal{U}}(\bm{y}^N) = \sum_{\bm{u}^N\in\mathcal{U}} P_{\bm{U}^N|\bm{Y}^N} \left(\bm{u}^N|\bm{y}^N\right),
\end{equation}
where $\mathcal{U}$ contains all valid input vectors $\bm{u}^N$ that satisfies the frozen constraints. Given a (partial) decoding codeword $\bm{a}^{(l),i}$ at the $l$-th path, the path probability $P_{\bm{U}^i|\bm{Y}^N}$ is associated with its \ac{PM} $\PM^{(l)}_{i}$ by \cite{Balats2015LLR}
\begin{equation}
	P_{\bm{U}^i|\bm{Y}^N} (\bm{a}^{(l),i}|\bm{y}^N) = e^{-\PM^{(l)}_{i}},
\end{equation}
which inspires us to compute $P_{\mathcal{U}}(\bm{y}^N)$ using the \acp{PM} output from \ac{SCL} decoder. However, it is challenging to access all valid path, especially with a realistic list size. 

\begin{figure}[!t]
    \centering
    \includegraphics[width=0.46\textwidth]{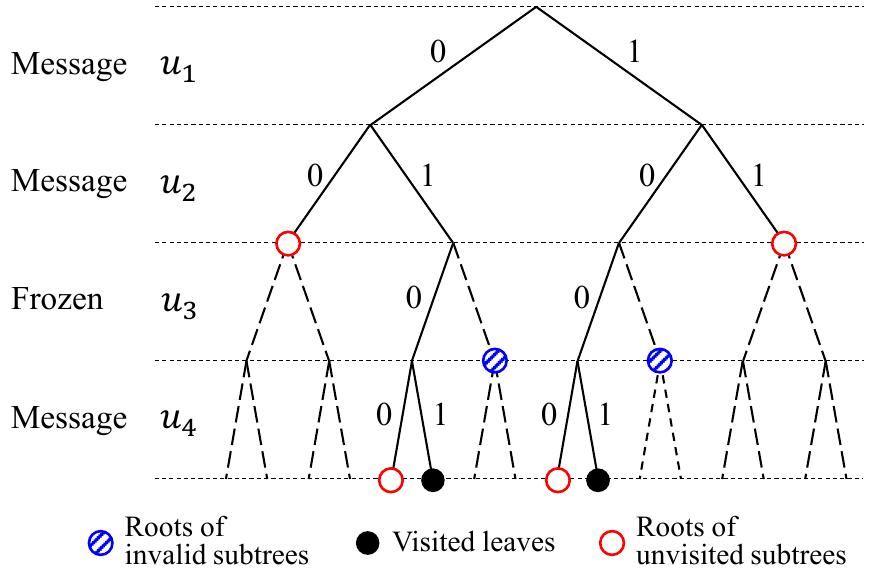}
    \caption{An example of the SCL decoding tree of a $(4,3)$ polar code with frozen bit $u_3=0$ and list size $L=2$. The whole tree consists of invalid subtrees rooted at $\mathcal{B} =\{(0,1,1), (1,0,1)\}$, visited leaves at $\mathcal{V}=\{(0,1,0,1), (1,0,0,1)\}$, and unvisited subtrees rooted at $\mathcal{W}=\{(0,0), (1,1), (0,1,0,0), (1,0,0,0)\}$.}
    \label{fig:DecTree}
	\vspace{-5pt}
\end{figure}

Hence, \cite{Yuan2024Near} proposed to approximate $P_{\mathcal{U}}(\bm{y}^N)$ by leveraging the \ac{SCL} decoding tree. An \ac{SCL} decoding tree, illustrated in Fig. \ref{fig:DecTree}, consists of three parts: leaves visited by \ac{SCL} decoding, unvisited valid subtrees, and invalid subtrees, where each node at the $i$-th level (root node is at the $0$-th level) corresponds to a possible decoding path $\bm{a}^i$, and each leaf thus represents a possible input vector $\bm{u}^N \in \{0, 1\}^N$. The unvisited subtrees and the invalid subtrees are pruned due to limitations on list size and conflict of frozen constraints, respectively. Let $\mathcal{V}$, $\mathcal{W}$, and $\mathcal{B}$ denote the sets of visited leaves, roots of unvisited subtrees, and roots of invalid subtrees, respectively. Then, the codebook probability $P_{\mathcal{U}}(\bm{y}^N)$ is approximated by \cite{Yuan2024Near}
\begin{equation}
\begin{aligned}
	P^*_{\mathcal{U}}(\bm{y}^N) &= \underbrace{\sum\nolimits_{\bm{u}^N\in \mathcal{V}} P_{\bm{U}^N|\bm{Y}^N}\left(\bm{u}^N|\bm{y}^N\right)}_{\text{(a) sum of prob. for all visited leaves}} \\
	&+ \underbrace{\sum\nolimits_{\bm{a}^i\in \mathcal{W}} 2^{-\left|\mathcal{F}^{(i:N)}\right|} P_{\bm{U}^i|\bm{Y}^N}\left(\bm{a}^i|\bm{y}^N\right)}_{\text{(b) approx. sum of prob. for all unvisited valid leaves}},
\end{aligned}
\label{eq:Pu_appr}
\end{equation}
where $\mathcal{F}^{(i:j)}$ contains frozen indices between $i$ and $j$, defined by $\mathcal{F}^{(i:j)} = \{k : k\in\mathcal{F}, i < k \leq j \}$. 

Now, the \ac{APP} \acp{LLR} $\ell_{\text{APP},i}$ are calculated by Eq. (\ref{eq:approx_app}) at the top of next page \cite{Yuan2024Near}, where $\mathcal{C}=\{\bm{c}^N: \bm{c}^N=\bm{u}^N\bm{G}_N,\forall \bm{u}^N\in\mathcal{U}\}$ and $\mathcal{V}_{c_i}^j=\{\bm{u}^N: c_i=j,\bm{c}^N=\bm{u}^N\bm{G}_N,\bm{u}^N\in\mathcal{V}\}$.

\begin{figure*}[!t]
\begin{equation}
\begin{aligned}
	\ell_{\text{APP},i} &\triangleq \ln\frac{P_{C_i|\bm{Y}^N}\left(0\,|\bm{y}^N\right) }{P_{C_i|\bm{Y}^N}\left(1\,|\bm{y}^N\right) } 
	= \ln \frac{\sum_{c_i=0, \bm{c}^N\in\,\mathcal{C}} P_{\bm{C}^N|\bm{Y}^N}\left(\bm{c}^N|\bm{y}^N\right)}
	{\sum_{c_i=1, \bm{c}^N\in\,\mathcal{C}} P_{\bm{C}^N|\bm{Y}^N}\left(\bm{c}^N|\bm{y}^N\right)} \\
	&\approx \ln \frac{\sum_{\bm{u}^N\in\mathcal{V}_{c_i}^0} P_{\bm{U}^N|\bm{Y}^N}\left(\bm{u}^N|\bm{y}^N\right) + \left(P_\mathcal{U}^*\left(\bm{y}^N\right) - \sum_{\bm{u}^N\in\mathcal{V}} P_{\bm{U}^N|\bm{Y}^N}\left(\bm{u}^N|\bm{y}^N\right)\right) \cdot P_{C|Y}\left(0\,|y_i\right)}
	{\sum_{\bm{u}^N\in\mathcal{V}_{c_i}^1} P_{\bm{U}^N|\bm{Y}^N}\left(\bm{u}^N|\bm{y}^N\right) + \left(P_\mathcal{U}^*\left(\bm{y}^N\right) - \sum_{\bm{u}^N\in\mathcal{V}} P_{\bm{U}^N|\bm{Y}^N}\left(\bm{u}^N|\bm{y}^N\right)\right) \cdot P_{C|Y}\left(1\,|y_i\right)}, i\in[\![N]\!].
\end{aligned}
\label{eq:approx_app}
\end{equation}
\hrulefill \vspace*{-5pt}
\end{figure*}

\section{Proposed SO-FSCL Decoding}
To reduce the decoding latency of \ac{SO-SCL}, we consider identifying four special nodes as introduced in \cite{Hashemi2017Fast} for \ac{FSCL} decoding: \ac{Rate0}, \ac{Rate1}, \ac{REP}, and \ac{SPC} nodes. Assuming that the indices in $\bm{u}^N$ of the $(N_s, K_s)$ sub-polar code underneath a special node start at $i_s$, the sub-codeword is then generated by $\bm{s}^{N_s} = \bm{u}_{[\![i_s, j_s]\!]} \bm{G}_{N_s}$, where $j_s=i_s+N_s-1$. We denote such a special node by $\mathbb{N}_{i_s}^{j_s}$ and the set of surviving nodes (paths) before decoding the node $\mathbb{N}_{i_s}^{j_s}$ by $\mathcal{V}_{i_s \minus 1}^\text{Node}$. Let $\mathcal{F}_s$ indicate the frozen positions of the sub-polar code.

The essence of \ac{FSCL} decoding is to directly obtain a list of estimates on sub-codeword $\bm{s}^{N_s}$ by exploiting the special properties of the sub-polar codes, instead of sequentially deciding $u_{i}$, $i\in[\![i_s, j_s]\!]$, as in \ac{SCL} decoding. As such, the \ac{PM} calculation in Eq. (\ref{eq:PM_u}) is no longer applicable. Equivalently, \ac{PM} can be calculated at the codeword side as \cite{Hashemi2016Fast}
\begin{equation}
	\PM^{(l)}_{j_s} = \PM^{(l)}_{i_s \minus 1} + \sum_{k=1}^{N_s} \ln\left(1 + e^{-(1-2\hat{s}^{(l)}_{k})\alpha^{(l)}_{k}}\right),
	\label{eq:PM_s}
\end{equation}
where $\hat{\bm{s}}^{(l)}$ is the estimated sub-codeword at the $l$-th path and $\bm{\alpha}^{(l)}$ is the internal \acp{LLR} passed to this node during \ac{SCL} decoding. The estimate $\hat{\bm{s}}^{(l)}$ can either serve as the internal result for the subsequent decoding, or be used to obtain an estimate of $\bm{u}_{[\![i_s, j_s]\!]}$ by polar transform $\hat{\bm{u}}_{[\![i_s, j_s]\!]}^{(l)} = \hat{\bm{s}}^{(l)} \bm{G}_{N_s}$.

Observing that the \ac{FSCL} decoder can output a list of candidate codewords like the conventional \ac{SCL} decoder, the key to soft output thus lies in the calculation of term (b) in Eq. (\ref{eq:Pu_appr}) for the considered special nodes. Moreover, the approximated codebook probability $P^*_{\mathcal{U}}(\bm{y}^N)$ is based on the assumption that $u_i$ is uniformly distributed for $i\in[\![N]\!]$ \cite{Yuan2024Near}, which implies that dynamic frozen bits are required. Therefore, we will further discuss the \ac{SO-SCL} decoding for dynamic frozen bits in Sec. \ref{sec:dyn_frz}.

\subsection{SO-FSCL Decoder for the Four Nodes}\label{sec:SO-FSCL}
For simplicity, we first assume all-zeros frozen bits. The fast decoding and soft information extraction for these nodes are described as follows.
\begin{figure*}[!t]
	\centering
    \includegraphics[width=0.99\textwidth]{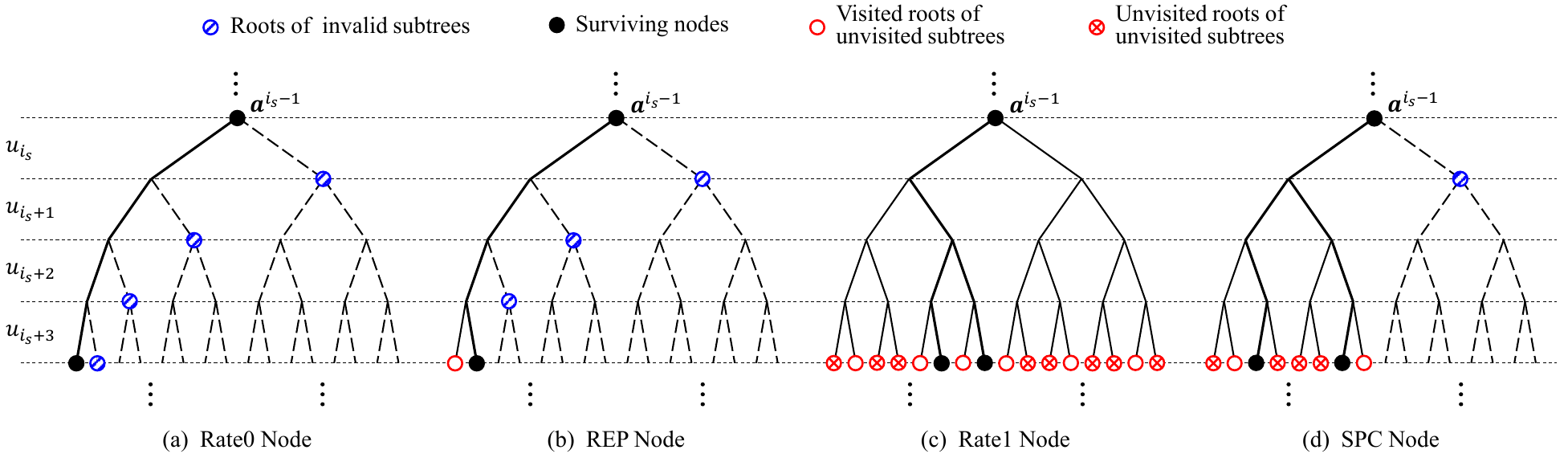}
    \caption{Examples of the partial \ac{FSCL} decoding tree for length-4 \ac{Rate0}, \ac{REP}, \ac{Rate1}, and \ac{SPC} nodes underneath a decoded path $\bm{a}^{i_s-1}$.}
    \label{fig:FSCLNode}
	\vspace{-5pt}
\end{figure*}
\subsubsection{Rate0 Node}
For a \ac{Rate0} Node, $\mathcal{F}_s = [\![N_s]\!]$. There is only one valid codeword, i.e., all-zeros codeword, as shown in Fig. \ref{fig:FSCLNode}(a). Therefore, no path splitting is required, and \acp{PM} for all paths are updated according to Eq. (\ref{eq:PM_s}).
\subsubsection{REP Node}
The \ac{REP} node is represented as $\mathcal{F}_s = [\![N_s-1]\!]$, which results in two valid codewords: all-zeros codeword and all-ones codeword. To decode this node, each path is split into two paths corresponding to these two codewords, and $L$ paths with the lowest \acp{PM} are retained. Fig. \ref{fig:FSCLNode}(b) provides an example of decoding a \ac{REP} node. After decoding a $\mathbb{N}_{i_s}^{j_s}$ \ac{REP} node, all valid nodes at the $j_s$-th level underneath nodes $\bm{a}^{i_s \minus 1} \in \mathcal{V}_{i_s \minus 1}^\text{Node}$ are visited, while the subtrees underneath some of them are pruned due to path selection. Let $\mathcal{W}_{i_s,j_s}^\text{REP}$ be the set containing roots of these pruned (unvisited) subtrees. Then, term (b) in Eq. (\ref{eq:Pu_appr}) is updated by
\begin{equation}
\begin{aligned}
	P^*_\mathcal{W}\left(\mathbb{N}_{i_s}^{j_s}, \text{REP}\right) &= \\
	&\hspace{-20pt} \sum_{\hspace{10pt} \bm{a}^{j_s} \in \mathcal{W}_{i_s,j_s}^\text{REP}} \hspace{-10pt} 2^{-\left|\mathcal{F}^{(j_s:N)}\right|} P_{\bm{U}^{j_s}|\bm{Y}^N}\left(\bm{a}^{j_s}|\bm{y}^N\right).
\end{aligned}
\label{eq:PW_REP}
\end{equation}
\subsubsection{Rate1 Node}
A \ac{Rate1} node contains no frozen bit, i.e., $\mathcal{F}_s = \emptyset$. Since it is impractical to traverse all $2^{N_s}$ valid codewords, \cite{Sarkis2016Fast} and \cite{Hashemi2017Fast} introduced a bit-flipping based approach for searching $L$ candidate codewords. Specifically, a \ac{ML} codeword is first obtained by performing hard decision on $\bm{\alpha}^{(l)}$ for each path. Then, flip the least reliable bit in \ac{ML} codewords to double the paths, and $L$ ones with the lowest \acp{PM} survive. Repeat this step by flipping from the least reliable bit to the most reliable one. The minimum required number of bit flips for \ac{Rate1} nodes is proved to be $\min(L-1,N_s)$, and more flips are redundant \cite{Hashemi2017Fast}. As illustrated in Fig. \ref{fig:FSCLNode}(c), although all nodes at the $j_s\text{-th}$ level underneath nodes $\bm{a}^{i_s \minus 1} \in \mathcal{V}_{i_s \minus 1}^\text{Node}$ are valid, the decoding algorithm for \ac{Rate1} nodes can only visit a portion of them. Nevertheless, observing that the equation
\begin{equation*}
	P_{\bm{U}^{i_s \minus 1}|\bm{Y}^N} \left(\bm{a}^{i_s \minus 1}|\bm{y}^N\right) = \hspace{-20pt} \sum_{\hspace{15pt} \bm{e}^{N_s}\in\{0,1\}^{N_s}} \hspace{-22pt}  P_{\bm{U}^{j_s}|\bm{Y}^N} \left([\bm{a}^{i_s \minus 1}, \bm{e}^{N_s}]|\bm{y}^N\right)
\end{equation*}
holds for all $\bm{a}^{i_s \minus 1} \in \mathcal{V}_{i_s \minus 1}^\text{Node}$, we update term (b) in Eq. (\ref{eq:Pu_appr}) by
\begin{equation}
\begin{aligned}
	&P^*_\mathcal{W}\left(\mathbb{N}_{i_s}^{j_s}, \text{Rate1}\right) = 2^{-\left|\mathcal{F}^{(j_s:N)}\right|} \times \\ 
	&\ \Big( \hspace{-16pt}\sum_{\hspace{15pt} \bm{a}^{i_s \minus 1} \in \mathcal{V}_{i_s \minus 1}^\text{Node}} \hspace{-20pt} P_{\bm{U}^{i_s \minus 1}|\bm{Y}^N} \left(\bm{a}^{i_s \minus 1}|\bm{y}^N\right) - 
	\hspace{-20pt} \sum_{\hspace{15pt} \bm{a}^{j_s} \in \mathcal{V}_{i_s,j_s}^\text{Rate1}} \hspace{-18pt} P_{\bm{U}^{j_s}|\bm{Y}^N}\left(\bm{a}^{j_s}|\bm{y}^N\right) \hspace{-3pt}\Big),
\end{aligned}
\label{eq:PW_Rate1}
\end{equation}
where the set $\mathcal{V}_{i_s,j_s}^\text{Rate1}$ contains surviving nodes after decoding a $\mathbb{N}_{i_s}^{j_s}$ \ac{Rate1} node.
\subsubsection{SPC Node}
For a \ac{SPC} node, $\mathcal{F}_s = \{1\}$, and the codeword satisfies that all bits sum up, modulo two, to zero. The decoding of \ac{SPC} nodes is similar to that of \ac{Rate1} nodes, with the only difference being that the candidate codewords should always satisfy the parity check. To this end, the least reliable bit of the \ac{ML} codeword is conditionally flipped to ensure the perpetual satisfaction of the parity check, and the generation of candidate codewords starts from the second least reliable bit. The required number of bit flips for \ac{SPC} node is $\min(L,N_s)$ \cite{Hashemi2017Fast}. Since the nodes at the $j_s\text{-th}$ level underneath nodes $[\bm{a}^{i_s \minus 1}, 0]$, $\bm{a}^{i_s \minus 1} \in \mathcal{V}_{i_s \minus 1}^\text{Node}$, are all valid but not all visited as shown in Fig. \ref{fig:FSCLNode}(d), We imitate decoding of \ac{Rate1} nodes to update term (b) in Eq. (\ref{eq:Pu_appr}) for a $\mathbb{N}_{i_s}^{j_s}$ \ac{SPC} node by \vspace{0cm}
\begin{equation}
\begin{aligned}
	&P^*_\mathcal{W}\left(\mathbb{N}_{i_s}^{j_s}, \text{SPC}\right) = 2^{-\left|\mathcal{F}^{(j_s:N)}\right|} \times \\ 
	&\ \Big( \hspace{-16pt}\sum_{\hspace{15pt} \bm{a}^{i_s \minus 1} \in \mathcal{V}_{i_s \minus 1}^\text{Node}} \hspace{-22pt} P_{\bm{U}^{i_s}|\bm{Y}^N} \hspace{-2pt} \left([\bm{a}^{i_s \minus 1}, 0]|\bm{y}^N\right) - 
	\hspace{-20pt} \sum_{\hspace{15pt} \bm{a}^{j_s} \in \mathcal{V}_{i_s,j_s}^\text{SPC}} \hspace{-20pt} P_{\bm{U}^{j_s}|\bm{Y}^N} \hspace{-2pt} \left(\bm{a}^{j_s}|\bm{y}^N\right) \hspace{-3pt}\Big),
\end{aligned}
\label{eq:PW_SPC}
\end{equation}
where the set $\mathcal{V}_{i_s,j_s}^\text{SPC}$ contains surviving nodes after decoding and the path probability $P_{\bm{U}^{i_s}|\bm{Y}^N}$ is obtained by performing an \ac{SCL} decoder for only decoding the frozen bit $u_{i_s}$.
\subsubsection{Calculation of the Bit-Wise APP LLRs}
Note that any polar code can be represented as a combination of several \ac{Rate0}, \ac{REP}, \ac{Rate1}, and \ac{SPC} sub-polar codes, since a polar code of length-2 must be one of these nodes. Let $\mathcal{N}$ be the set of special nodes that constitute a polar code. We approximate the codebook probability $P_{\mathcal{U}}(\bm{y}^N)$ for proposed \ac{SO-FSCL} decoding as follows
\begin{equation}
\begin{aligned}
	P^*_{\mathcal{U}}(\bm{y}^N) =& \sum\nolimits_{\bm{u}^N\in \mathcal{V}} P_{\bm{U}^N|\bm{Y}^N}\left(\bm{u}^N|\bm{y}^N\right) +\\
	&\sum\nolimits_{\mathbb{N}_{i_s}^{j_s}\in\mathcal{N}} P^*_\mathcal{W}\left(\mathbb{N}_{i_s}^{j_s}, \mathcal{T}(\mathbb{N}_{i_s}^{j_s})\right),
\end{aligned}
\label{eq:Pu_apprF}
\end{equation}
where $\mathcal{T}(\cdot)$ returns the type of a special node. Specifically, $P^*_\mathcal{W}(\,\cdot\,, \text{Rate0}) = 0$ as no path is split, while $P^*_\mathcal{W}(\,\cdot\,, \,\cdot\,)$ is calculated according to Eq. (\ref{eq:PW_REP}), (\ref{eq:PW_Rate1}), and (\ref{eq:PW_SPC}) for \ac{REP}, \ac{Rate1}, and \ac{SPC} nodes, respectively. Thus, the proposed \ac{SO-FSCL} decoder outputs bit-wise \ac{APP} \acp{LLR} by substituting Eq. (\ref{eq:Pu_apprF}) into Eq. (\ref{eq:approx_app}).

\subsection{Compatibility with Dynamic Frozen Bits}\label{sec:dyn_frz}
When decoding a $\mathbb{N}_{i_s}^{j_s}$ node, the dynamic frozen bits $\hat{u}^{(l)}_{i+i_s-1}$, $i \in \mathcal{F}_s$, for each path is required to be determined according to $\hat{\bm{u}}^{(l)}_{[\![i+i_s-2]\!]}$ beforehand. Let $\tilde{\bm{u}}^{N_s} = \bm{u}_{[\![i_s, j_s]\!]}$ for clarity. By observing the structure of these four nodes, we can write $\tilde{\bm{u}}^{N_s}$ into a cascading form of $\tilde{\bm{u}}_{\mathcal{F}_s}$ and $\tilde{\bm{u}}_{\mathcal{I}_s}$, i.e. $\tilde{\bm{u}}^{N_s} = [\tilde{\bm{u}}_{\mathcal{F}_s}, \tilde{\bm{u}}_{\mathcal{I}_s}]$, where $\mathcal{I}_s = [\![N_s]\!] \cap \mathcal{F}_s^C$. Hence, the sub-codeword $\bm{s}^{N_s}$ is represented as
\begin{equation}
	\bm{s}^{N_s} = \tilde{\bm{u}}^{N_s} \bm{G}_{N_s} = [\tilde{\bm{u}}_{\mathcal{F}_s}, \bm{0}^{|\mathcal{I}_s|}] \bm{G}_{N_s} + [\bm{0}^{|\mathcal{F}_s|}, \tilde{\bm{u}}_{\mathcal{I}_s}] \bm{G}_{N_s},
\end{equation}
where we denote the term $[\tilde{\bm{u}}_{\mathcal{F}_s}, \bm{0}^{|\mathcal{I}_s|}] \bm{G}_{N_s}$ by $\bm{s}_F^{N_s}$. If we treat $\bm{s}^{N_s} \minus \bm{s}_F^{N_s}$ as the sub-codeword under all-zero frozen bits assumption, the internal \acp{LLR} passed to this node are then modified to 
\begin{equation}
	\tilde{\alpha}^{(l)}_{k} = (1-2s_{F,k})\alpha^{(l)}_{k},
	\label{eq:MdLLR}
\end{equation}
for all $k\in[\![N_s]\!]$ based on the definition of \ac{LLR}. Thus, we can apply the node decoding introduced in Sec. \ref{sec:SO-FSCL} to generate estimates of $\bm{s}^{N_s} \minus \bm{s}_F^{N_s}$, and obtain estimates of $\bm{s}^{N_s}$ under dynamic frozen conditions.

Generally, we should perform one matrix multiplication or polar encoding to compute $\bm{s}_F^{N_s}$, causing undesired additional latency. To avoid the matrix multiplication, we propose to set only partial frozen bits in the special nodes to be dynamic. Specifically, we choose the first $F_d$ frozen bits (if have) to be dynamic, where $F_d$ should be small enough so that we can find the corresponding $\bm{s}_F^{N_s}$ from $2^{F_d}$ possible sub-codewords via a look-up table. As such, we can immediately obtain $\bm{s}_F^{N_s}$ according to $\tilde{\bm{u}}_{[\![F_d]\!]}$.

\section{Decoding Latency Analysis}
The decoding latency can be evaluated by the required number of time steps to decode a special node. We adopt the following assumptions used in \cite{Condo2018Generalized,Hashemi2017Fast,Ardakani2019Fast,Zheng2021Threshold} for analyzing decoding latency: 1) we assume that there is no resource limitation for operations that can be executed in parallel, 2) basic operation of real numbers and check-node operation require one time step, 3) hard decisions, bit operations, and sign operations can be carried out instantaneously, 4) it takes one time step to obtain the \ac{ML} codeword of a \ac{SPC} node, and 5) path splitting, the sorting of \acp{PM} and the selection of the most probable paths consume one time step. Furthermore, we assume that a single dynamic frozen bit can be computed immediately, while the computation of a dynamic frozen bit sequence requires one time step.

Since \ac{SO-SCL} relies on the conventional \ac{SCL} decoder, it takes $2N+K-2$ time steps to generate the hard-output for a $(N,K)$ polar code \cite{Hashemi2017Fast}. Meanwhile, the update of term (b) in Eq. (\ref{eq:Pu_appr}) can be done in parallel. After completing \ac{SCL} decoding, \ac{SO-SCL} decoder can immediately obtain the approximation in Eq. (\ref{eq:Pu_appr}) and consume an additional time step to calculate \ac{APP} \ac{LLR} for each bit according to Eq. (\ref{eq:approx_app}).

\begin{table}[!t]
	\caption{Required Number of Time Steps to Decoding Different Nodes of Length $N_s$ and List Size $L$}
	\centering
	\begin{tabular}{@{}lcccc@{}}
		\toprule
		Algorithms & Rate0 & REP & Rate1 & SPC \\ \midrule
		FSCL\cite{Hashemi2017Fast} & 1 & 2 & \hspace{-5pt}$\min(L,N_s\plus 1)$\hspace{-5pt} & $\min(L,N_s)$ \vspace{4pt}\\
		SO-SCL\cite{Yuan2024Near} & \hspace{-5pt}$2N_s\minus2$\hspace{-5pt} & $2N_s\minus1$ & $3N_s\minus2$ & $3N_s\minus3$ \vspace{1pt}\\
		SO-FSCL & 2 & 3 & \hspace{-5pt}$\min(L,N_s\plus 1)$\hspace{-5pt} & \begin{tabular}[c]{@{}c@{}} $\max(\log_2 N_s,$ \\ $\min(L, N_s))$\end{tabular} \\ 
		\bottomrule
	\end{tabular}
	\label{tab:ts}
\end{table}

Our proposed \ac{SO-FSCL} decoder performs hard decoding of these considered nodes utilizing algorithms in \cite{Hashemi2017Fast}, which require $1$, $2$, $\min(L-1, N_s)+1$, and $\min(L, N_s)$ time steps to decode \ac{Rate0}, \ac{REP}, \ac{Rate1}, and \ac{SPC} nodes, respectively. For \ac{Rate0} and \ac{REP} nodes, \ac{SO-FSCL} decoder takes one time step for calculating dynamic frozen bits and modifying \acp{LLR} as in Eq. (\ref{eq:MdLLR}) beforehand, while we can easily obtain the modified \acp{LLR} for \ac{Rate1} and \ac{SPC} nodes, since there is at most one dynamic frozen bit and all operations are binary. After generating the sub-codeword for each node, \ac{SO-FSCL} decoder calculates $P^*_\mathcal{W}(\,\cdot\,, \,\cdot\,)$ within a time step, but this process can be executed in parallel with the subsequent decoding. Moreover, to calculate Eq. (\ref{eq:PW_SPC}) for \ac{SPC} node, \ac{SO-FSCL} decoder needs to consume $\log_2 N_s$ time steps to obtain the path probability $P_{\bm{U}^{i_s}|\bm{Y}^N}$ while performing hard decoding, which results in a decoding latency of $\max\left(\log_2 N_s, \min(L, N_s)\right)$ time steps for \ac{SPC} node decoding. Like \ac{SO-SCL} decoder, \ac{SO-FSCL} decoder finally estimates the \ac{APP} \acp{LLR} in one clock cycle. The decoding latency of the proposed \ac{SO-FSCL} decoder is summarized in Table \ref{tab:ts}.

\section{Numerical Results}
In this section, we evaluate the performance of the proposed \ac{SO-FSCL} decoder in terms of decoding latency, soft output, and application in \ac{MIMO} systems. We adopt 5G polar codes and generate dynamic frozen bits, as in \cite{Yuan2024Near}, by
\begin{equation}
	u_i = u_{i-2} \oplus u_{i-3} \oplus u_{i-5} \oplus u_{i-6}, i\in\mathcal{F}, i>6.
\end{equation}
The value $F_d$ is set to 3 in our simulation.

We first count the required time steps of \ac{SO-FSCL} decoding for polar codes with different code lengths and code rates, as shown in Table \ref{tab:simu_ts}. It is observed that the proposed \ac{SO-FSCL} decoder can save at least 76\% of the time steps with respect to the \ac{SO-SCL} decoder. However, compared to \ac{FSCL} decoder, \ac{SO-FSCL} decoder needs to generate soft output at the cost of a little decoding latency. Note that the reduction in decoding latency attributed to node-based decoding is related only to the structure of polar codes and is independent of the channel conditions.

\begin{table}[!t]
	\caption{Required Number of Time Steps to Decoding Different Polar Nodes with List Size $L=4$}
	\centering
	\begin{tabular}{@{}cccc@{}}
		\toprule
		$(N,K)$ & SO-SCL\cite{Yuan2024Near} & FSCL\cite{Hashemi2017Fast} & SO-FSCL \\ \midrule
		$(128,85)$ & 595 & 121 & 137 \\
		$(512, 256)$ & 1278 & 232 & 259 \\
		$(1024, 512)$ & 2558 & 402 & 450 \\ 
		\bottomrule
	\end{tabular}
	\label{tab:simu_ts}
\end{table}

\begin{figure}[!t]
	\centering
	\includegraphics[width=0.46\textwidth]{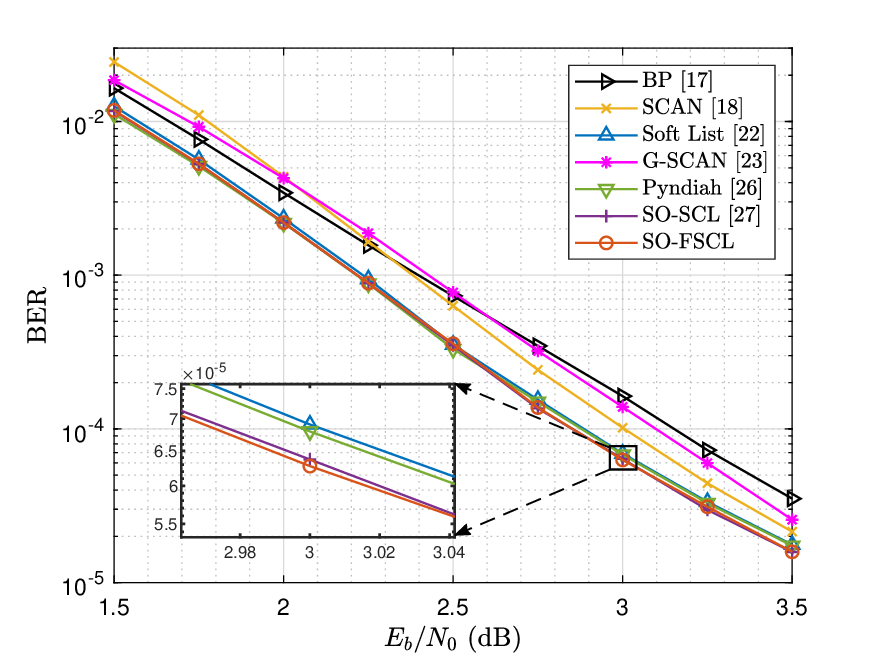}
	\caption{BER performance of various soft-output polar decoders for a $(512, 256)$ polar code.}
    \label{fig:BER}
\end{figure}

Since an \ac{APP} decoder estimates each bit according to $\hat{c}_i = \arg\max_{a\in\{0,1\}} P_{C_i|\bm{Y}^N}(a|\bm{y}^N)$, we assess the \ac{BER} performance by performing hard decisions on \ac{APP} \acp{LLR} output by difference soft-output polar decoders, as displayed in Fig. \ref{fig:BER}. The number of inner iterations for \ac{BP} decoder \cite{Arikan2008Perform} and \ac{SCAN} decoder \cite{Fayyaz2014Low} is $I_{\text{BP}}=50$ and $I_{\text{SCAN}}=20$, respectively, while the list size for soft list decoder \cite{Xiang2020Soft}, G-SCAN decoder \cite{Egilmez2022soft}, Pyndiah's approximation \cite{Pyndiah1998Near}, \ac{SO-SCL} \cite{Yuan2024Near}, and the proposed \ac{SO-FSCL} decoder is $L=2$. We observe that despite the utilization of fast decoding and partially dynamic frozen of bits, \ac{SO-FSCL} decoder shows no performance loss compared with \ac{SO-SCL} decoder and outperforms other soft-output decoders.

\begin{figure*}[t]
    \centering
    \captionsetup[subfloat]{justification=centering}
    \subfloat[$N=256$, $K=85$, and CRC-$6$.]{
        \includegraphics[width=0.33\textwidth]{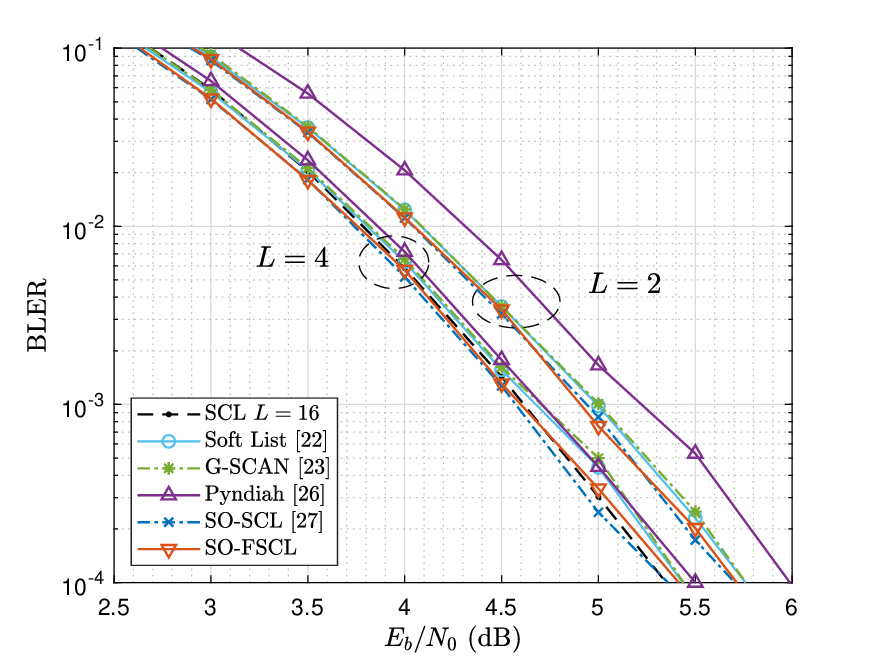}} \hspace{-0.5cm}
    \subfloat[$N=512$, $K=256$, and CRC-$11$.]{
        \includegraphics[width=0.33\textwidth]{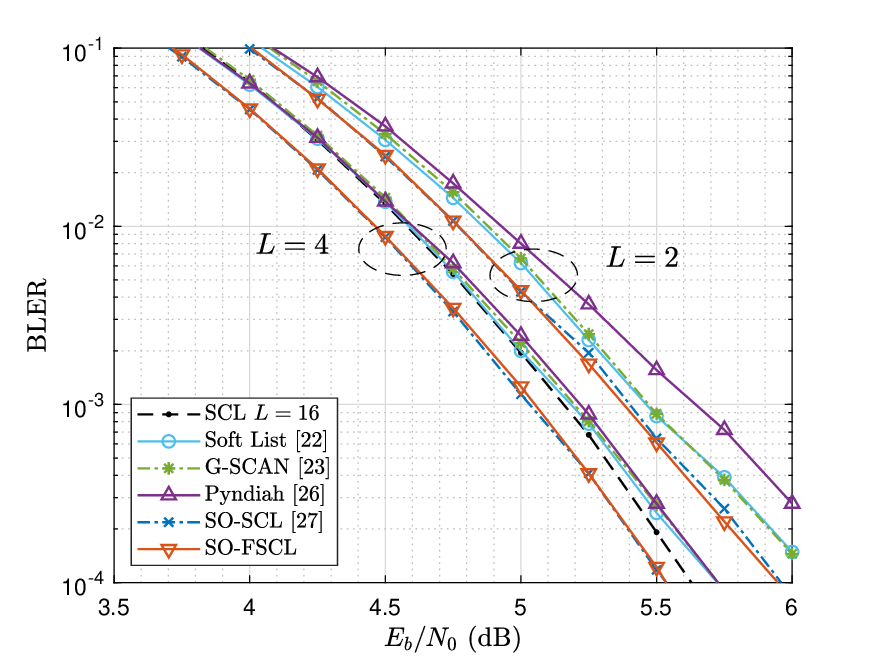}} \hspace{-0.5cm}
    \subfloat[$N=1024$, $K=512$, and CRC-$11$.]{
        \includegraphics[width=0.33\textwidth]{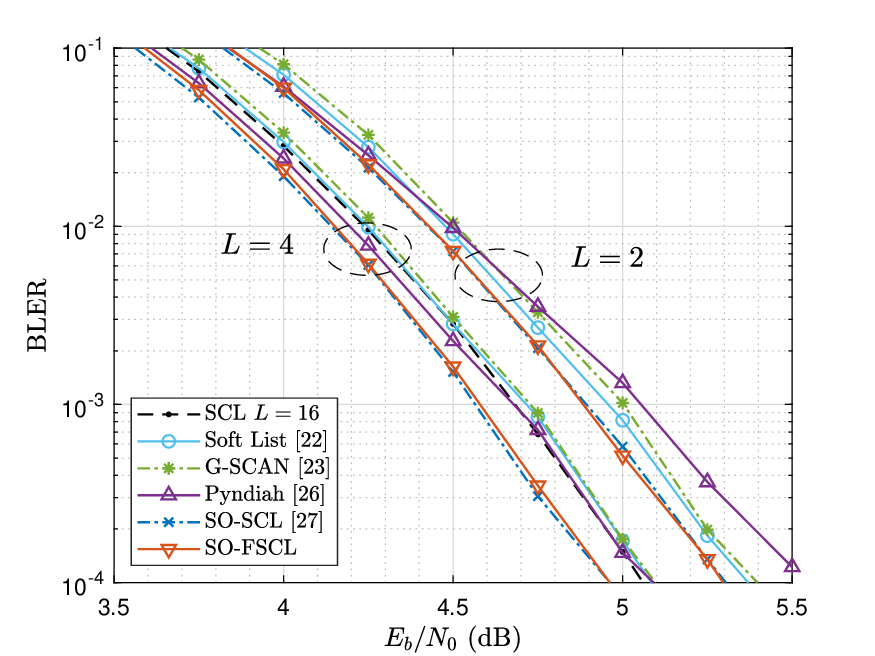}}
    \caption{BLER performance of various soft-output polar decoders over QPSK-input $2\times 2$ MIMO channel for different polar codes.}
    \label{fig:BLER}
\end{figure*}

To further illustrate the performance of proposed \ac{SO-FSCL} decoder, we apply these decoders to $2\times 2$ \ac{MIMO} systems with iterative decoding \cite{Hochwald2003Ach} and \ac{QPSK} input. Each element of the channel matrix follows an independent and identically distributed Gaussian distribution with zero mean and unit variance. At the receiver, the channel state information is assumed to be known and a maximum a posteriori detector is used. Fig. \ref{fig:BLER} shows the \ac{BLER} performance of \ac{SO-FSCL} decoders with a maximum of $5$ iterations for different polar codes. The results of hard-output \ac{SCL} decoder with $L=16$ and no iteration are also provided as a benchmark. Analogously to \text{Fig. \ref{fig:BER}}, \ac{SO-FSCL} decoder exhibits negligible performance loss compared to \ac{SO-SCL} decoder. As the list size $L$ grows, the gains of \ac{SO-FSCL} and \ac{SO-SCL} decoders increase since the estimated \ac{APP} \acp{LLR} are related to $L$. Furthermore, our simulation results also imply that the proposed partially dynamic with $F_d=3$ will not incur performance loss compared to the fully dynamic frozen bits of the \ac{SO-SCL} decoding.

\section{Conclusion}
We proposed a \ac{SO-FSCL} decoder by identifying \ac{Rate0}, \ac{Rate1}, \ac{REP}, and \ac{SPC} nodes. Underneath these nodes, the calculation of codebook probability $P_{\mathcal{U}}(\bm{y}^N)$ is investigated. Considering the dynamic frozen bits, we set the frozen bits to be partially dynamic to facilitate fast decoding. Simulation results show that the proposed \ac{SO-FSCL} decoder can significantly reduce the decoding latency without loss of decoding performance compared to \ac{SO-SCL} decoder.

\section*{Acknowledgment}
The work of Y. Wu is supported in part by the Fundamental Research Funds for the Central Universities, National Natural Science Foundation of China (NSFC) under Grant 62122052 and 62071289, 111 project BP0719010, and STCSM 22DZ2229005.

\bibliographystyle{IEEEtran}
\bibliography{refer}

\end{document}